\documentclass[proof]{WileyASNA-v1}

\articletype{Article Type}%

\received{26 April 2016}
\revised{6 June 2016}
\accepted{6 June 2016}

\newcommand{\HI}{H{\,\scriptsize I}}

\newcommand{\kms}{$\,$km$\,$s$^{-1}$}
\newcommand{\WHz}{$\,$W$\,$Hz$^{-1}$}

\newcommand{\msun}{{$M_\odot$}}
\newcommand{\msunyr}{{$M_\odot$ yr$^{-1}$}}

\begin{document}

\title{The impact of young radio jets traced by cold molecular gas}

\author[1,2]{Raffaella Morganti*}

\author[1,2]{Tom Oosterloo}

\author[1,2]{Suma Murthy}

\author[3]{Clive Tadhunter}

\authormark{Morganti \textsc{et al}}

\address[1]{ASTRON, the Netherlands Institute for Radio Astronomy, Oude Hoogeveensedijk 4, 7991 PD, Dwingeloo, The Netherlands.}

\address[2]{Kapteyn Astronomical Institute, University of Groningen, Postbus 800,
9700 AV Groningen, The Netherlands}

\address[3]{Department of Physics and Astronomy, University of Sheffield, Sheffield, S7 3RH, UK}


\corres{*R. Morganti, \email{morganti@astron.nl}}


\abstract{
Ranging from a few pc to hundreds of kpc in size, radio jets have, during their evolution, an impact on their gaseous environment on a large range of scales. 
While their effect on larger scales is well established,  it is now becoming clear that they can also strongly affect the interstellar medium (ISM) inside the host galaxy. 
Particularly important is the initial phase ($<10^6$ yr) of the evolution of the radio jet, when they expand into the inner few kpc of the host galaxy.
Here we report on  results obtained for a representative group of young radio galaxies using the cold molecular gas as a tracer of  jet-ISM interactions.
The sensitivity and high spatial resolution of ALMA and NOEMA are ideal to study the details of this process.

In many objects we find  massive molecular outflows   driven by the plasma jet, even in low-power radio sources. However, the observed outflows   are limited to the circumnuclear regions and only a small fraction of the ISM is leaving the galaxy. Beyond this region, the impact of the jet  seems to change. Fast outflows are replaced by a milder expansion  driven by the expanding cocoon created by the  jet-ISM interaction, resulting in dispersing and heating the ISM.  
These findings are in line with  predictions from simulations of jets interacting with a clumpy medium and suggest a more complex view of the impact of AGN than  presently implemented in cosmological simulations. 
}

\jnlcitation{\cname{
\author{Morganti R. et al.}, 
} 
(\cyear{2021}), 
\ctitle{The impact of young radio jets traced by cold molecular gas}, \cjournal{Astron. Nachr.}, \cvol{2021;XX:X--X}.}

\maketitle

\begin{figure*}[t]
\centerline{\includegraphics[scale=0.20]{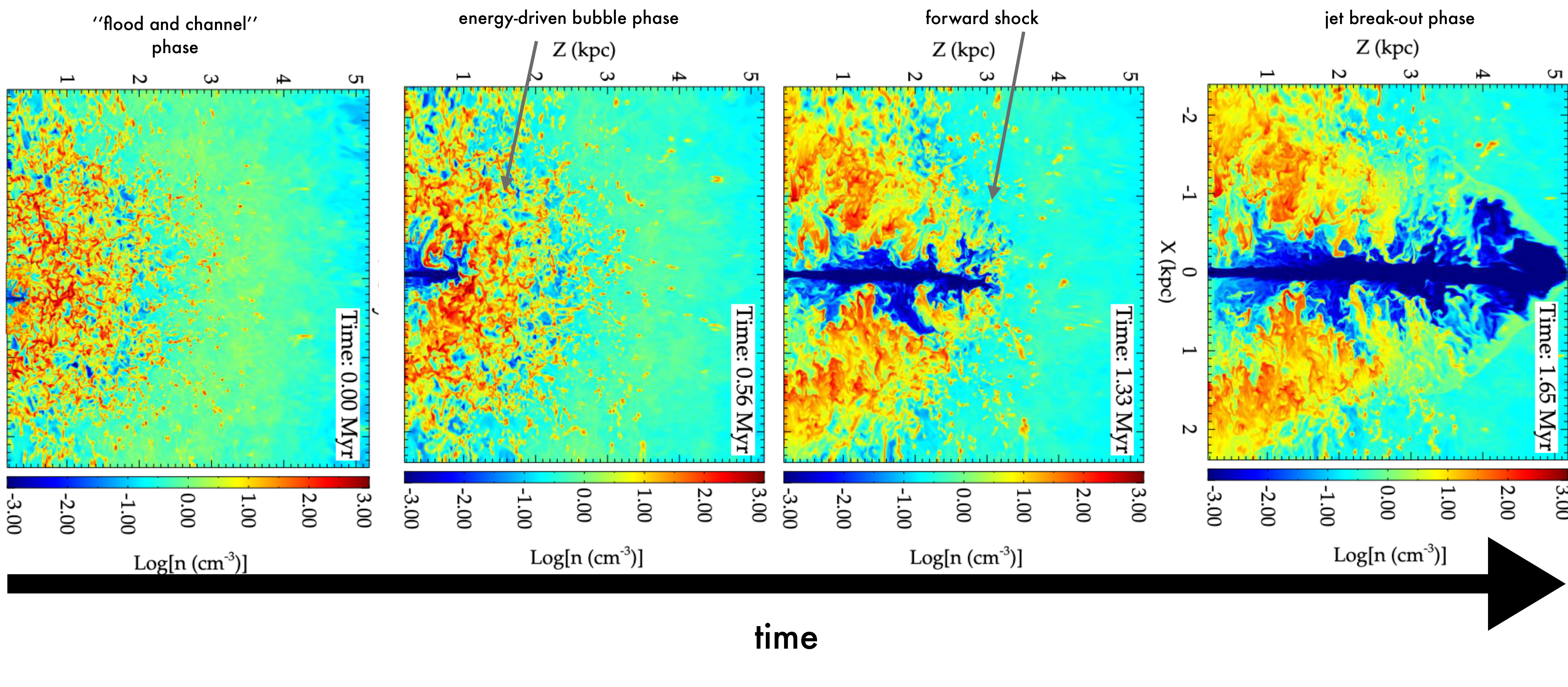}}
\caption{Simulations of a radio jet expanding in a clumpy medium in the inner 5 kpc of the host galaxy (images taken from \citealt{Mukherjee16}). The different phases and their impact are marked following \cite{Sutherland07}, see text. 
\label{fig1}}
\end{figure*}

\section{Impact of (young) radio jets from small to large scales}\label{sec1}

The impact on the surrounding medium  of the energy released by super-massive black holes in their active phase (AGN) is an important ingredient of models of galaxy evolution. However, this effect, known as AGN feedback, has proven to be more complex than often implemented in cosmological simulations (e.g.\ \citealt{Weinberger17,Zinger20}). This is due to the large parameter space that determines the type and magnitude of feedback  and its evolution during the life of the host galaxy.  Radio jets are known to be relevant for AGN feedback. However, it is often thought that their main role is limited to preventing the cooling of gas of the large-scale inter-galactic and intra-cluster medium, the so-called {\sl maintenance mode} (e.g.\ McNamara et al.\ 2012). Although this is some of the best evidence for AGN feedback, recent results show that the impact of jets starts at  an early stage  in the inner, circumnuclear regions of the host galaxy. This highlights the important role played by young radio jets (i.e.\ with ages $\ll\,10^6$ yr), which can, for example, drive fast gas outflows. 
Compact Steep Spectrum (CSS) and GigaHertz Peaked Spectrum (GPS) sources represent candidate young radio sources \citep{ODea21} and are ideal targets  to study the details of this impact.

The presence of jet-driven outflows traced by warm ionised gas is known since a long time  (e.g.\ \citealt{Whittle88,Capetti99} and refs therein) as outflows of warm ionised gas (with velocities up to more than 2000 \kms) are found to be common in young radio galaxies (e.g.\ \citealt{Holt08,Shih13}). 
However, the associated mass outflow rates of these outflows are modest (typically below 1 \msunyr) and the impact - in terms of the ratio between the kinetic energy of the outflow and the bolometric or Eddington luminosity - is limited (see e.g.\  \citealt{Holt11} and refs therein).

However, two main aspects have recently revamped the interest in the relevance of radio jets for feedback:
the discovery of young radio galaxies with massive jet-driven  outflows of cold (\HI\ and molecular) gas which are significantly more massive than those of warm ionised gas (see \citealt{Morganti18} for a review on the \HI),
and the predictions from more realistic numerical simulations of the impact of jet-ISM interactions (e.g.\ \citealt{Mukherjee18a}).
Most importantly, these numerical simulations show that a large parameter space needs to be explored in order to fully quantify the impact of the jet.
These results form the basis of the work presented here.

The general relevance of AGN-driven outflows of cold molecular gas is motivated by the finding that they appear to  carry most of the mass of the overall outflow. Their mass outflow rate is typically found to be much higher than what is associated with warm, ionised gas (see \citealt{Veilleux20} for an overview).  
The presence of cold gas in AGN-driven outflows is surprising at first sight, but has been explained by the very rapid cooling that dense, shocked gas can experience (e.g.\ \citealt{Richings18,Mukherjee18b}). 

\section{The observed sample and the predictions from the simulations}\label{sec2}
Numerical simulations of jet-ISM interactions (\citealt{Mukherjee18a} and refs therein) show that the impact of the jet may depend on a number of parameters, like jet power, age of the jet, and the orientation between the jet and the distribution of the ISM.  
Because of this, we have embarked on a project aimed to study the properties of the cold molecular gas in a representative sample of  gas-rich young radio galaxies covering this parameter space. The observations are obtained with high enough spatial resolution to resolve the distribution and kinematics of the gas across the radio emission. 
For our project,  we have used ALMA and, more recently, the NOrthern Extended Millimeter Array (NOEMA). Here we present a brief summary of some of the results obtained so far.

We have observed  7  CSS and GPS sources in CO(1-0) or CO(2-1) at  spatial resolutions ranging between 0.2 and 1.5 arcsec, enabling to spatially resolve the distribution of the CO along the radio continuum emission. For two more targets the observations are in progress. 
Only for two objects observations of two or more CO transitions are available. 
The sources were mainly selected from \cite{Holt08} and \cite{Gereb15} so that also  information about the other component of the cold gas, \HI,  is available. Based on their \HI\ properties, two more sources were included (IC~5063 and PKS~1718-64).  The sample is, therefore, selected to include some of the best cases where to study the process of jet-ISM interaction. 
Furthermore, the objects were selected to cover a variety of properties of  young radio galaxies. They range from extremely small and young sources (like PKS~1718--64: 2 pc in size and a $\sim 10^2$ yr age of the radio activity; \citealt{Maccagni18}) to older and larger, kpc-scale sources (like PKS~0023--26: $\sim 4$ kpc in size and $\sim 10^6$ yr old; \citealt{Morganti21}).  Furthermore, the sample  includes radio-quiet, low radio power radio sources ($<10^{24}$ \WHz) as well as very powerful ones ($> 10^{27}$ \WHz).

To guide the interpretation of the observations, we use the state-of-the-art hydrodynamic simulations of  jet-ISM interactions, of  \cite{Sutherland07,Wagner12} and \cite{Mukherjee18a}. An important result of these simulations is that the jet couples very strongly to the ISM if propagating in a clumpy medium. Furthermore, as shown in Fig.\ \ref{fig1}, the simulations suggest  four phases in the evolution of the jets, each providing a different type of impact on the surrounding ISM: 1) an initial ``flood and channel'' phase, where the expansion of the jet strongly depends on the interaction with high-pressure clumps of gas; 2) as consequence of this interaction, the formation of a spherical, energy-driven bubble phase starts;  3) a subsequent, rapid phase where the jet breaks free from the last obstructing dense clouds and,  4) a classical phase, where the jet propagates to large scales in a momentum-dominated fashion. The first phase is where we expect  fast and massive outflows, while in the final phases the effect of the jet may act more in preventing the ISM/IGM gas to cool. The duration and details of these phases depend on jet power. Interestingly, simulations predict that even low-power jets are able to impact the medium, as their phase of interaction  with the ISM will last longer  and, therefore, they will inject their energy for a longer time (see \citealt{Mukherjee18a}).

\begin{figure}
	\centerline{\includegraphics[width=75mm]{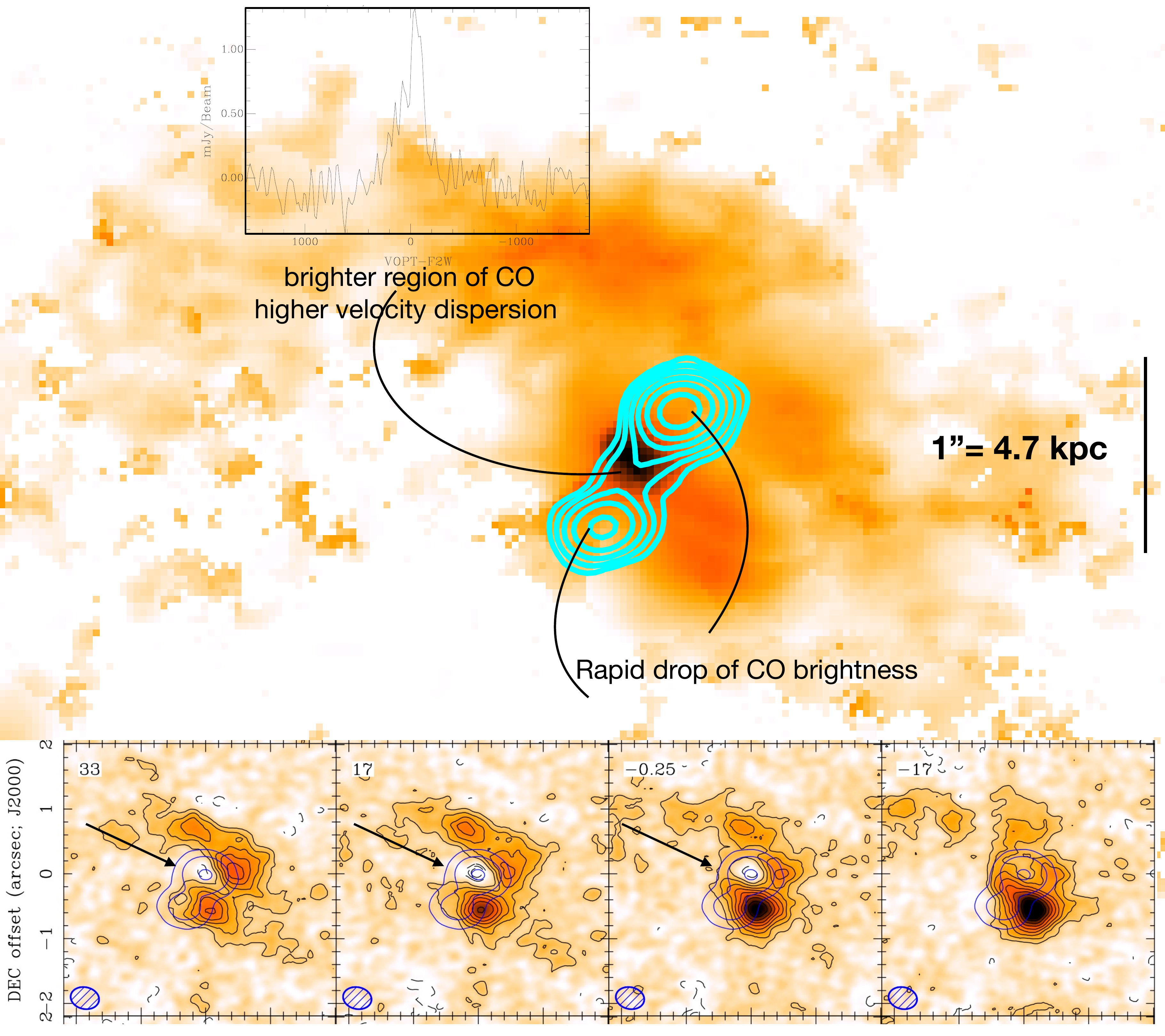}}
	\caption{{Top -} Distribution of the molecular gas (CO(2-1) from ALMA) of PKS~0023--26. The contours of the 3-mm continuum emission are indicated in cyan. The locations of the most relevant features are indicated. {Bottom -} Four channel maps showing the distribution of the gas at different velocities (33, 17 ,--0.25, --27 \kms\ w.r.t.\ the systemic velocity). Some of the regions/velocities where the molecular gas is seen to wrap around the radio  lobes are indicated for clarity. \label{fig3}}
\end{figure}

\section{Result so far: covering a large parameter space}\label{sec3}

Despite the limited sample studied so far, we already have a number of interesting insights.

\begin{itemize}

\item The only case where we could identify infalling clouds of molecular gas, likely connected with the fuelling of the AGN, is PKS~1718--64 \citep{Maccagni18}. Interestingly, this is the smallest (and youngest) radio source in the sample. With only 2~pc in size, the radio plasma may not have been able yet to affect the surrounding medium. The detection of infalling cloud appears to be consistent with a fuelling process, as predicted by the chaotic cold accretion (e.g.\ \citealt{Gaspari17}) and similar to the case of  PKS~2322-123 in Abell 2597 \citep{Tremblay16}. 

\item Molecular outflows are detected in a number of objects. 
In the best studied case, IC~5063 \citep{Morganti15,Oosterloo17}, we have
detected kinematically disturbed molecular gas along the full extent of the radio source (about 1 kpc). The comparison between  observations and simulations shows that the radio jets can produce the observed distribution and kinematics of the   molecular gas (\citealt{Mukherjee18b}).

\item In the dust-obscured, young and powerful radio galaxy PKS~1549--79, we detect, using ALMA, one of the most massive molecular gas outflows ($\sim 650$ \msunyr) likely driven by the radio jet in the process of clearing its way out from the enshrouding dense gas \citep{Oosterloo19}. However, the outflow is limited to the inner 200 pc of the galaxy, despite the presence of a powerful jet and also a powerful quasar AGN. A circumnuclear disc of $M_{\rm H_2} = 2.6 \times  10^8$ \msun\ is observed and appears to co-exist with the outflow. This means that, unless the gas in the disc is replenished, on a time scale of $\sim 10^5$ yr the AGN would be able to destroy the central disc of molecular gas and deplete the central region of this gas.

\item Outflows are also found to be driven by low-power jets. The NOEMA CO(1-0) observations of B2 0258+35 ($L\rm_{1.4 GHz}=2.1 \times 10\rm^{23}~\rm W~\rm Hz^{-1}$; Murthy et al.\ submitted) show a spectacular example of this and further confirms the predictions from  simulations. We detect a highly turbulent molecular circumnuclear structure,  where  a fast (FWHM $\sim 350$ km s$^{-1}$) jet-driven outflow of $\sim  2.6 \times 10^6$ \msun\ is observed. This outflow comprises of $\sim 75$\% of the total gas in the nuclear region. Also in this case, the jet will deplete the kpc-scale molecular gas reservoir on a relatively short time scale (i.e.\ within $2 \times 10^6$ yr).

\begin{figure}
	\centerline{\includegraphics[scale=0.40]{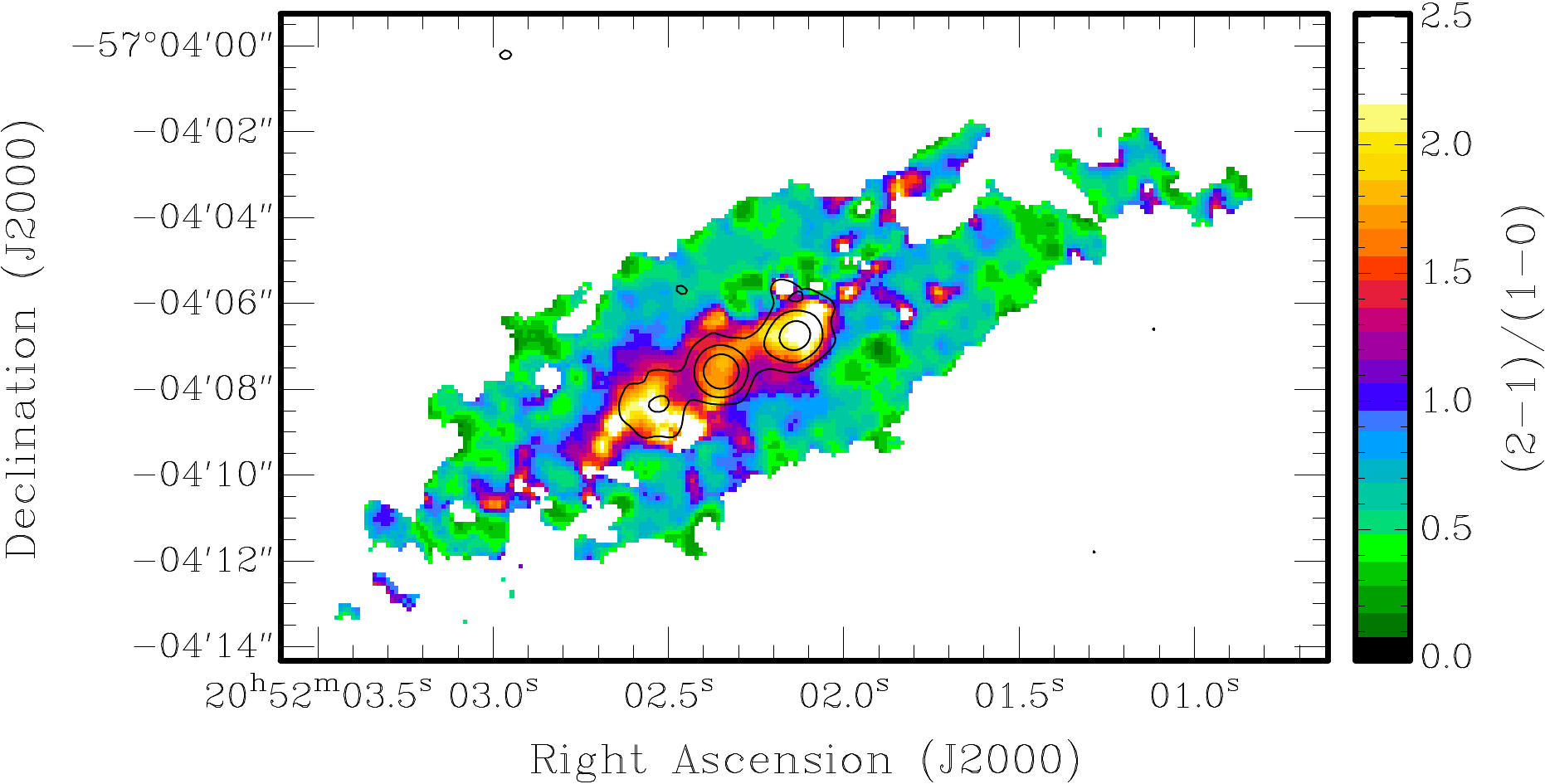}}
	\caption{Line ratio CO(2-1)/CO(1-0) of IC~5063 with overlaid the radio continuum contours. A sharp difference is visible between the region co-spatial with the radio jet and  where the gas has a quiescent kinematics
	(from \citealt{Oosterloo17}). \label{fig2}}
\end{figure}

\item We find mass molecular gas outflow rates  ranging from tens to a few hundred \msunyr, but only a relatively small fraction of the gas (at most $\sim 10$\%) leaves the galaxy: most of the gas will rain back in a  {\sl fountain-like effect}. Thus, our work suggests that molecular outflows, even if present, cannot be  the only effect responsible for the AGN feedback required in cosmological simulations. 

\item Interestingly, in the most extended ($\sim 4$ kpc) and powerful radio source of the sample (PKS~0023--26, which also hosts a quasar AGN), only a mild outflow is observed limited to the sub-kpc region. Instead, on kpc scales, the molecular gas is distributed predominantly around the radio lobes forming a bubble-like structure (see Fig.\ \ref{fig3} and \citealt{Morganti21}). 
This bubble is likely driven by the expansion of the cocoon created by the  jet-ISM interaction,  pushing aside the preexisting molecular gas and resulting in dispersing and heating the molecular clouds. This is more similar to the ``maintenance'' phase,  preventing the gas from cooling, known to happen on even larger scales.
Thus, these ALMA observations suggest that the mode of coupling between radio jets and the ISM could change as the jet expands and that, already on galaxy scales, the impact of the AGN is not limited to outflows. 

\item Finally, also the physical conditions of the molecular gas are affected by the impact of the (radio) AGN.
This has been seen using multiple CO transitions in the cases of IC~5063 (see Fig.\  \ref{fig2}) and PKS~1549--79, see \cite{Oosterloo17} and \cite{Oosterloo19} respectively.
The line ratios of the CO transitions indicate that the gas in the region affected by the interaction with the radio plasma has different excitation and/or optical thickness, as result of the impact from strong shocks. Instead, where the gas has  quiescent kinematics, the line ratios are similar to what observed in normal gas discs in galaxies.

\end{itemize}

\section{Conclusions and future work}\label{sec5}

The results obtained, even if limited to a small number of cases, show the impact of radio jets on the surrounding molecular gas, demonstrating the relevance for feedback. Confirming  predictions of the simulation for the importance of low-power radio sources is also important: these sources are relatively common in massive galaxies \citep{Best05} and, therefore, they can provide an important population of AGN for feedback. These results will be expanded to more objects (two more sources are in the process of being observed) to better sample how the impact changes with the properties of the jet and to improve the comparison with the predictions of the simulations.

\section*{Author Biography}
\begin{biography}
{\includegraphics[width=60pt]{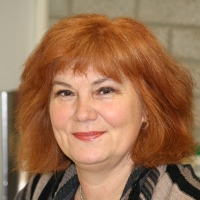}}
{\textbf{Raffaella Morganti} is senior astronomer at the Netherlands Institute of Radio Astronomy and affiliated to the university of Groningen as professor. She works on radio AGN and, in particular, on the gas content and properties of these objects in relation to their evolution and life-cycle. This has been the focus of her ERC-AdG RadioLife (see also \url{http://astron.nl/~morganti}).}
\end{biography}
\end{document}